\documentclass[showpacs,amsmath,amssymb,prl,superscriptaddress,floatfix, preprint]{revtex4}
\usepackage{graphicx}
\usepackage{dcolumn}
\usepackage{bm}

\begin{document}

\title{Tunable Fermi acceleration in the driven elliptical billiard}
\author{F. Lenz}
\email[]{lenz@physi.uni-heidelberg.de}
\affiliation{Physikalisches Institut,  Universit\"at Heidelberg,
Philosophenweg 12, 69120 Heidelberg, Germany}
\author{F.K. Diakonos}
\affiliation{Department of Physics, University of Athens, GR-15771 Athens, Greece}
\author{P. Schmelcher}
\affiliation{Physikalisches Institut,  Universit\"at Heidelberg,
Philosophenweg 12, 69120 Heidelberg, Germany}

\affiliation{Theoretische Chemie, Physikalisch-Chemisches Institut,
Universit\"at Heidelberg, Im Neuenheimer Feld 229,
 69120 Heidelberg, Germany}
\date{\today}

\begin{abstract}
We explore the dynamical evolution of an ensemble of
non-interacting particles propagating freely in an elliptical
billiard with harmonically driven boundaries. The existence of
Fermi acceleration is shown thereby refuting the established
assumption that smoothly driven billiards whose static
counterparts are integrable do not exhibit acceleration dynamics.
The underlying mechanism based on intermittent phases of laminar
and stochastic behavior of the strongly correlated angular
momentum and velocity motion is identified and studied with
varying parameters. The diffusion process in velocity space is
shown to be anomalous and we find that the corresponding
characteristic exponent depends monotonically on the breathing
amplitude of the billiard boundaries. Thus it is possible to tune
the acceleration law in a straightforwardly controllable manner.
\end{abstract}
\maketitle

Fermi acceleration was first proposed in 1949 \cite{Fermi:1949} to explain the high energies of cosmic ray particles  interacting with a time-dependent magnetic field (for a review see
\cite{Blandford:1987}). Nowadays, Fermi acceleration is investigated in a variety of  systems belonging to different areas of physics, such as astrophysics
\cite{Veltri:2004,Kobayakawa:2002,Malkov:1998}, plasma physics
\cite{Michalek:1999,Milovanov:2001}, atom optics
\cite{Saif:1998,Steane:1995} and has even been used for the interpretation of
experimental results in atomic physics \cite{Lanzano:1999}. In the context of
dynamical system theory, Fermi acceleration is defined as the unlimited growth
of the energy of a particle moving in a time-dependent potential. For the
experimental realization of Fermi acceleration in the laboratory, driven, i.e. time-dependent
billiards are the most promising devices. A particle propagating in these
systems gains or looses energy whenever it hits the billiards' moving boundary.
An infinite sequence of such collisional events can in principle lead to
Fermi acceleration of the scattered particle. Driven billiards can be realized experimentally in various ways, such as (superconducting) mesoscopic cavities (see ref. \cite{Stockmann:1999, Richter:2001} and refs. therein) or atom optical setups employing acousto-optical scanners for the (time-dependent) deflection of correspondingly detuned laser beams \cite{Milner:2001}.

The simplest realization of a driven billiard is the well-known Fermi-Ulam model
(FUM) \cite{Ulam:1961,Lieberman:1972,Lichtenberg:1992} which consists of an
ensemble of non-interacting freely moving particles in one dimension bouncing
between an oscillating and a fixed wall. The FUM and its variants have been the
subject of extensive theoretical (see Ref.~\cite{Lieberman:1972} and references
therein) and experimental
\cite{Kowalik:1988,Warr:1996,Celaschi:1987} studies. It has been proven
\cite{Lichtenberg:1992} that  a necessary condition for the development of Fermi acceleration in the
FUM  is the non-smooth (non-differentiable) dependence
of the velocity of the moving wall on time (this includes especially randomized dynamical systems).  An
alternative scenario for the occurrence of Fermi acceleration in $1D$ systems
is a bouncer model \cite{Pustilnikov:1983} with an
additional nonlinear time-independent term in the potential.

In contrast to the $1D$ case, driven billiards with a higher dimensional configuration space pose many open questions. The lack
of the possibility to visualize  phase space in terms of
Poincar\'{e} surfaces of section complicates the analysis of e.g. $2D$
time-dependent billiards \cite{Koiller:1995}. Despite this fact, several studies
of Fermi acceleration in these systems
\cite{Koiller:1996,Loskutov:1999,Loskutov:2000,de Carvalho:2006} have been
performed. These investigations provide us with the conclusion that a sufficient
condition for the occurrence of Fermi acceleration in a $2D$ smoothly driven
billiard is the existence of a chaotic part in the phase space of the corresponding
time-independent system, obtained by assuming static boundaries (see the LRA conjecture in ref.
\cite{Loskutov:1999}). This is supported by the absence of Fermi
acceleration in the smoothly oscillating circular billiard where the
corresponding static system is integrable \cite{Kamphorst:1999}. On the other
hand the static Lorentz gas possesses a predominantly chaotic phase space and its harmonically driven counterpart exhibits Fermi acceleration \cite{Loskutov:2000,Karlis:2006}. Similar results were obtained  for the
oscillating stadium-like \cite{Loskutov:2002}, oval \cite{Leonel:2006} as well
as annular billiard \cite{de Carvalho:2006}, each of them possessing chaotic portions with respect to the phase space of the corresponding time-independent system.

To elucidate the above-described problem of the existence of Fermi acceleration for driven billiards whose time-independent counterparts are integrable, we investigate here the elliptical harmonically driven billiard.  According to the existing literature the ellipse
should not exhibit Fermi acceleration (see e.g. ref. \cite{Koiller:1996}). We show that, opposite to what is expected in the literature, Fermi
acceleration occurs for ensembles of particles propagating in a
breathing elliptical billiard even for very small oscillation amplitudes. The driving causes a layer of unstable motion around the separatrix of the elliptical billiard leading to large fluctuations of the velocities of the particles as they cross it.  These  fluctuations, which increase with time, lead to an anomalous
diffusion in velocity space and an unlimited growth of the mean kinetic energy
of the multiple scattered particles. The anomalous behavior  of
the velocity diffusion process is due to the intermittent character of the dynamics
attributed to the  instability of the fixed points in the regime of librators. We determine the
dependence of the exponent of the  acceleration law on the
amplitude of the  oscillation of the boundary leading to the conclusion that the
smoothly driven elliptical billiard represents a tunable source of Fermi acceleration.

\begin{figure}[htbp]
\centerline{\includegraphics[width=\columnwidth]{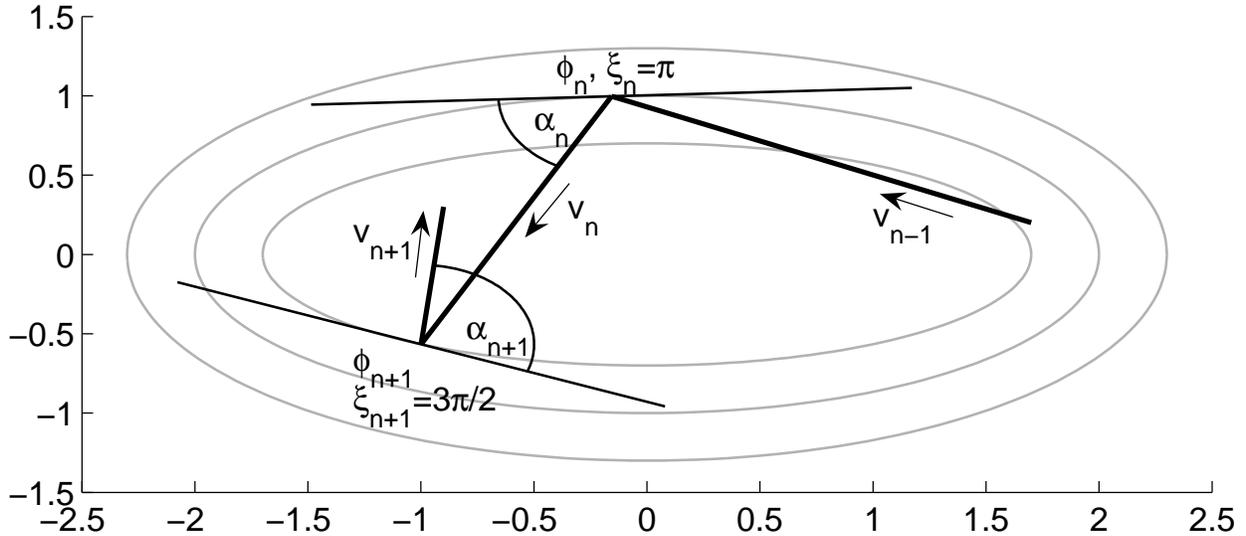}}
\caption{Schematic picture of the setup of the elliptical
billiard.} \label{fig:fig1}
\end{figure}

Let us specify our setup (for a more detailed
description see ref. \cite{Lenz:2007}).
The static ellipse is integrable: besides the energy $E$ the
product of the angular momenta $F$ about the two foci of the ellipse
\cite{Berry:1982} is conserved.
\begin{equation}\label{eq:F}
    F(\varphi,p)= \frac{p^2 (1+ (1-\varepsilon^2) \cot^2
\varphi)-\varepsilon^2}{1+ (1-\varepsilon^2) \cot^2 \varphi -
    \varepsilon^2},
\end{equation}
where $\epsilon$ is the numerical eccentricity. The variables $p=\cos
\alpha$ and $\varphi$ are defined at the elliptical boundary of the billiard:
$\alpha$ is the angle between the tangent and the trajectory of the
colliding particle while $\varphi$ is the azimuthal angle, see eq. \eqref{eq:EllBoundary} and Fig. \ref{fig:fig1}. The phase space is globally divided by the
separatrix ($F=0$), associated with two hyperbolic fixed points, into rotators
($F~>~0$) and librators $F~<~0$. Two elliptic fixed points are
located at the global minimum $F_{min}$ of $F$.  The driving law of the elliptic boundary is:
\begin{equation}\label{eq:EllBoundary}
 \left( \begin{array}{c} x(t) \\ y(t) \\ \end{array}\right)=  \left (\begin{array}{c} (A_0 + C \sin (\omega t))
\cos \varphi \\
   (B_0 + C \sin (\omega t)) \sin \varphi\\ \end{array}\right )
\end{equation}
where $t$ is time, $(x(t),y(t))$ is a point on the boundary, $\varphi$ is
a $2\pi$-periodic parameter, $C>0$ is the driving
amplitude, $\omega$ is the frequency of oscillation and $A_0$, $B_0$ are the
equilibrium values of the long and the short half-diameter, respectively. The
phase of oscillation at $t=0$ is set to zero in  the following.
In order to reduce the number of parameters to be varied in our investigations, we  fix
$\omega=1$, $A_0=2$ and $B_0=1$ (arbitrary units). The scattering dynamics in the billiard is described by an implicit $4D$
map specifying the sequence of the collisional events on the boundary of
the ellipse. A convenient choice of the variables of this map is
$(\varphi_n,p_n,\vert \vec{v}_n \vert,\xi_n)$, where $\vert \vec{v}_n \vert$
is the magnitude of particles' velocity after the collision and $\xi_n$ is
the phase of the boundary oscillation at the $n$-th collision. The implicit equations defining this map are:
\begin{subequations}
\label{eq:Mapping}
\begin{equation}\label{eq:ImplTime}
    \left ( \frac{v_n^x (t_{n+1}-t_n) + x_n}{A_0 + C \sin (\omega t_{n+1})}
\right )^2 + \left (\frac{v_n^y (t_{n+1}-t_n) + y_n}{B_0 + C \sin (\omega t_{n+1})} \right )^2 -1 = 0,
\end{equation}
\begin{equation}\label{eq:MapPoint}
 \bm{x}_{n+1}=\bm{x}_n+\bm{v}_n(t_{n+1}-t_n)
\end{equation}
\begin{equation}\label{eq:MapVelocity}
 \bm{v}_{n+1} = \bm{v}_n -2 \left [ \hat{ \bm{n}}_{n+1} \cdot
(\bm{v}_n -
\bm{u}_{n+1})\right ] \cdot \hat{ \bm{n}}_{n+1}
\end{equation}
\end{subequations}
where the smallest $t_{n+1}>t_n$ that solves \eqref{eq:ImplTime} has to be
taken and $\xi_{n+1} = t_{n+1} \bmod 2\pi$. $\bm{x}_n=(x_n,y_n)$ is the $n$th collision point.
$\varphi_{n+1}$ can be obtained by  inverting
\eqref{eq:EllBoundary}. In eq. \eqref{eq:MapVelocity}, $\bm{u}_{n+1}$ is the boundary velocity and $\hat{ \bm{n}}_{n+1}$ the normal vector of the collisional event occurring at time $t_{n+1}$ and
position $\bm{x}_{n+1}$. When iterating the mapping \eqref{eq:Mapping} numerically, solving \eqref{eq:ImplTime} requires the major computational effort in terms of CPU-time, even when applying indispensable advanced bracketing techniques. When the driving is applied, both $E$ and $F$ are no longer conserved
quantities. However, they are still very valuable for the description of the dynamics in the presence of the driving \eqref{eq:EllBoundary},
 since at every time instant the elliptical form of the boundary is preserved.

\begin{figure}[htbp]
\centerline{\includegraphics[width=\columnwidth]{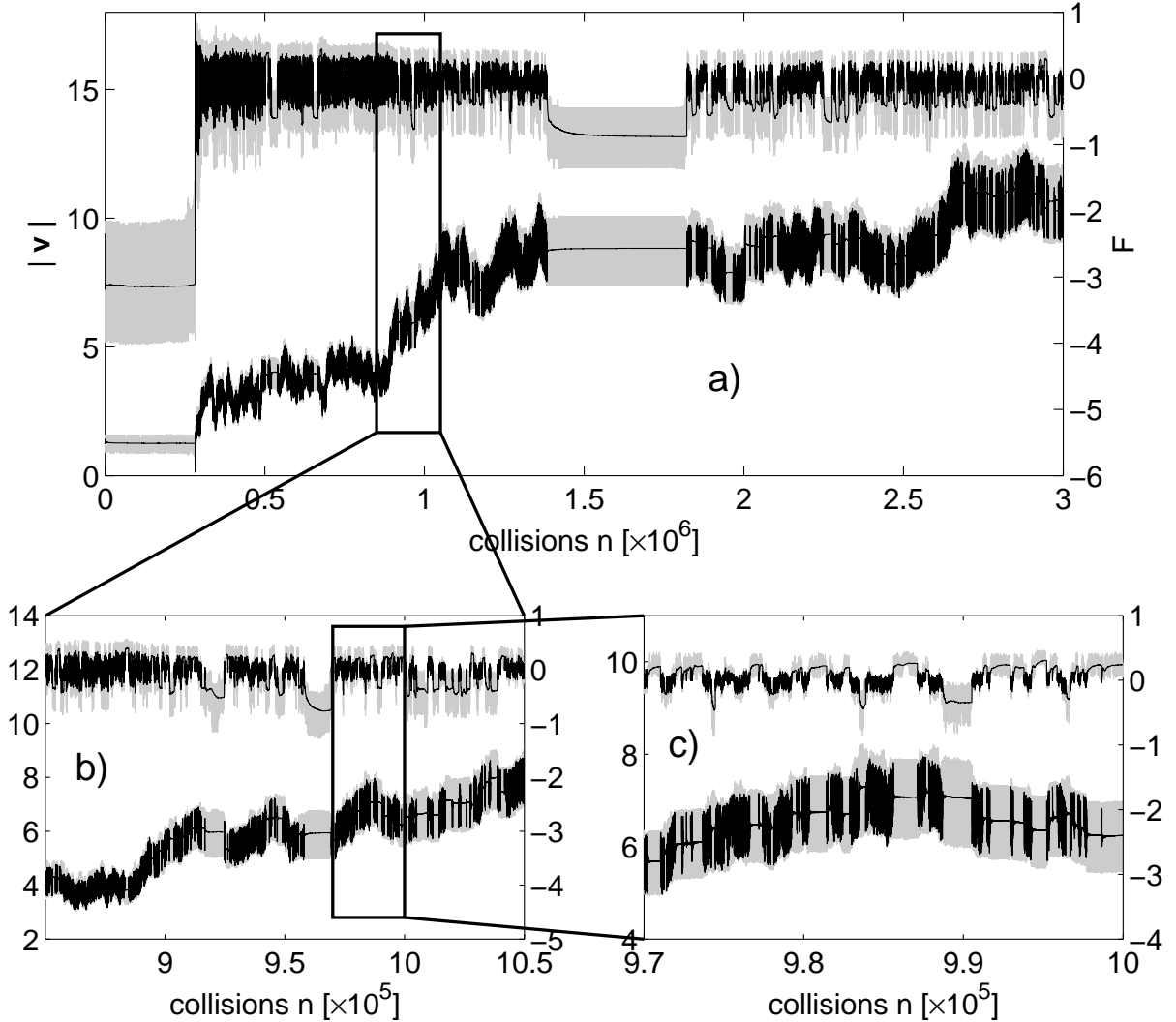}}
\caption{The functions $F(n)$ (upper curve, light gray) and $\vert
\vec{v} \vert(n)$ (lower curve, light gray) for a single
trajectory for $C=0.2$ in arbitrary units. The solid black lines
indicate the cumulative mean in each case. The two successive
magnifications demonstrate that the corresponding fluctuations
occur on many scales.} \label{fig:fig2}
\end{figure}

Before discussing the ensemble averaged properties of the time-dependent system,
 it is illuminating to analyze a typical trajectory being initially on a librator orbit ($C=0.2$). Fig.~\ref{fig:fig2} shows the evolution of $F$ (upper curve, light gray) and $\vert \vec{v} \vert$ (lower curve, light gray) respectively as a function of the number of
collisions $n$. According to Fig.~\ref{fig:fig2}a, $F(n)$ alternates between periods of regular (laminar phases) oscillations (intervals $[0, \,\,2.8\cdot10^5]$ and $[1.4\cdot 10^6, \,\,1.75\cdot 10^6]$) and periods of irregular fluctuations (turbulent phases). During these laminar phases, $F(n)$ never crosses the $F=0$ line, whereas during the turbulent phases $F(n)$ remains essentially within the zone $[F_{min},F_{max}]= [-1,0.4]$, repeatedly crossing the $F=0$ line associated with the separatrix (the separatrix is defined by $F=0$, independent of $t$). From Fig.~\ref{fig:fig2}b and \ref{fig:fig2}c we see that this structure of laminar and turbulent phases exists on different scales of $n$. Whenever $F(n)$ is in a laminar phase, $\vert \vec{v} \vert$ oscillates around a fixed central value, whereas during the turbulent $F(n)$ periods, $\vert \vec{v} \vert(n)$ starts to develop intervals with strong, irregular fluctuations leading to a sudden increase or decrease of its
central value. In fact, the increasing parts of the trajectory $\vert \vec{v} \vert(n)$ prevail such that a net
 increase of the velocity for longer times can be observed.

\begin{figure}[htbp]
\centerline{\includegraphics[width=\columnwidth]{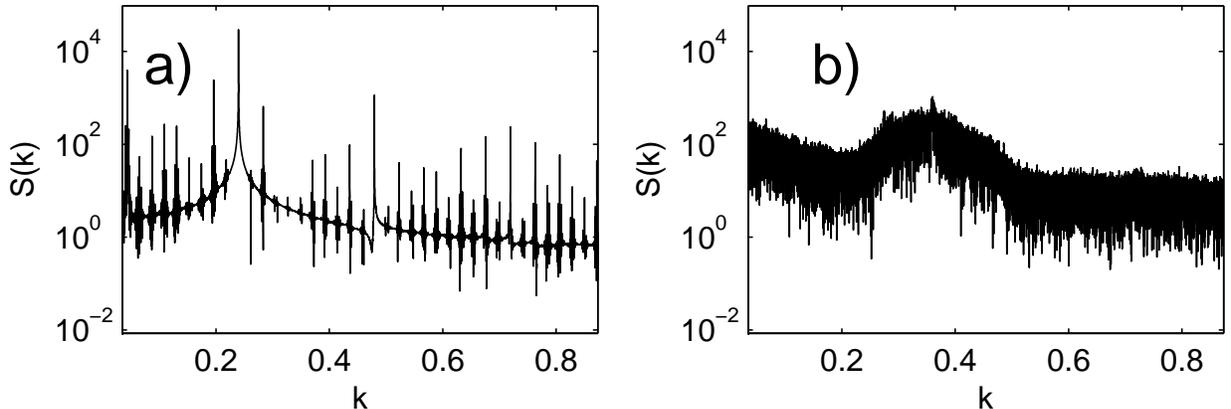}}
\caption{Power spectrum $S(k)$ of a typical laminar (a) and
turbulent phase (b) in arbitrary units.} \label{fig:fig3}
\end{figure}

The above-observed behavior indicates two important dynamical
properties of the driven system: (i) the region around the separatrix is characterized by
stochastic dynamics and (ii) the trajectories $F(n)$ and $\vert \vec{v} \vert(n)$  are
strongly correlated (see Fig.~\ref{fig:fig2}). The enhanced stochasticity of the dynamics around $F
\approx 0$ is also verified by the appearance of the corresponding power-spectrum $S(k) = \frac{2\pi}{N}\left \vert  \sum_{n=1}^{N} F(n)e^{\frac{-2\pi i(k-1)(n-1)}{N}}\right \vert$
presented in Fig.~\ref{fig:fig3}. Clearly, the continuous part of the spectrum is dominating in
this case (Fig.~\ref{fig:fig3}b). On the contrary, in  regions of large negative $F$ values
the power spectrum, shown in Fig.~\ref{fig:fig3}a, is dominated by isolated peaks while the
continuous background is much less pronounced. Furthermore, the correlated behavior of
$\vert \vec{v} \vert$ and $F$ suggests that an acceleration mechanism may be associated
with the crossing of the $F=0$ line.
To elucidate this,  it is necessary to integrate out
regular as well as stochastic fluctuations of the laminar phase occurring on
small time scales. For such an averaging procedure to be successful one has to detect
first efficiently the intervals of laminar evolution. A rough estimation of
these intervals is obtained by analyzing the timeseries $F(n)$. As `working intervals' (to calculate the cumulative mean) resembling the laminar phases, we identify the dynamics between two successive zero crossings of $F$. Then,
in each such interval $j$, we calculate the cumulative mean
$\bar{F}(n)_j=\frac{1}{n} \sum_{i=1}^n F^{(j)}_i$ and
$|\bar{\vec{v}}|(n)_j=\frac{1}{n} \sum_{i=1}^n \vert \vec{v}^{(j)}_i \vert$ as a
function of $n$. Here $n$ takes values $n=1,..,L_j$ with $L_j$ being the number
of collisions between the $j-1$-th and the $j$-th zero crossing of $F$. The
results for $\bar{F}_j$ and $|\bar{\vec{v}}|_j$ of the above-considered trajectory are
shown as solid black lines in
Fig.~\ref{fig:fig2}. The similarity of the  dynamics of the cumulative mean of $F$ with
a typical trajectory showing intermittency \cite{Procaccia:1983} is
remarkable. Additionally, the one-to-one correspondence between periods of stochastic behavior
(bursts) of $\bar{F}_j$ and periods of fluctuations of $|\bar{\vec{v}}|_j$ is obvious.
The intermittent character of the $F$-dynamics goes along with a specific distribution of the laminar lengths $L_j$, i.e. the intervals
between successive zeros of $F$, which according to the theory of intermittent
maps \cite{Schuster:1989} should obey a power-law. Indeed, we find asymptotically (for large values of $L_j$)
a power-law behavior of the distribution with a characteristic exponent of $\eta = -2.00 \pm 0.05$.

\begin{figure}[htbp]
\centerline{\includegraphics[width=\columnwidth]{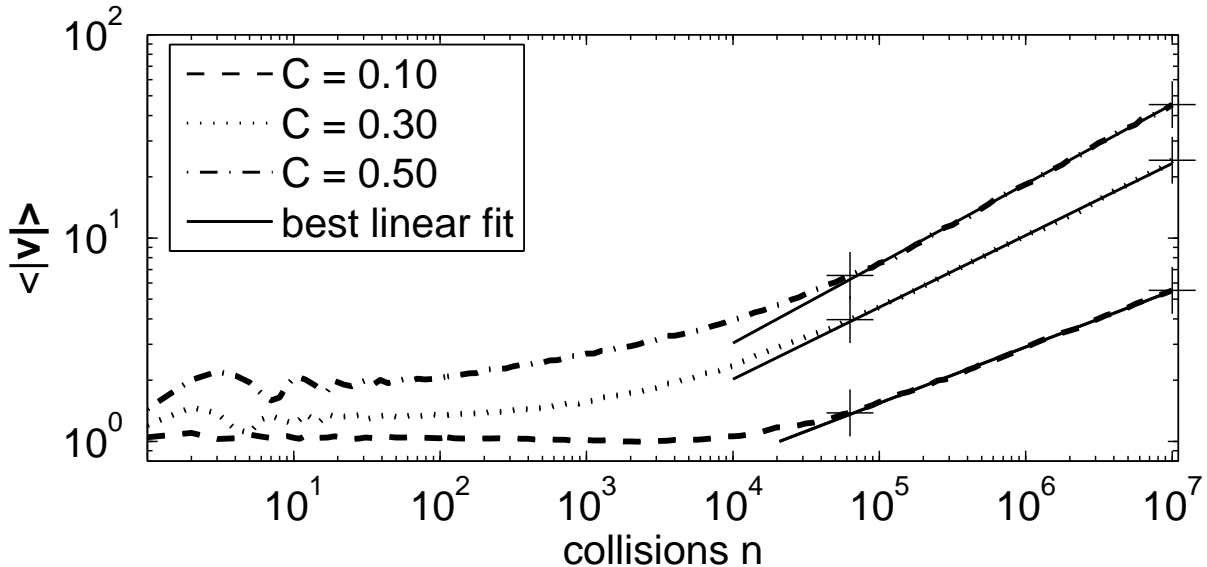}}
\caption{The function $\langle \vert \vec{v} \vert \rangle(n)$
(arbitrary units) for an ensemble of $10^3$ trajectories for
different values of the driving amplitude $C$, together with the
best linear fit (between the two crosses).} \label{fig:fig4}
\end{figure}

Let us now investigate whether Fermi acceleration occurs in the driven ellipse
by examining $\langle \vert \vec{v} \vert \rangle(n)$, where $\langle~.~\rangle$
denotes ensemble averaging. In Fig.~\ref{fig:fig4} we show the results of our simulations
obtained using $1000$ trajectories with initial values of $\alpha$ and $\varphi$
uniformly distributed in $[0,\pi]$ and $[0,2 \pi]$, respectively, for different
driving amplitudes $C$ (see also Table~I). Each trajectory is propagated $10^7$
collisions and possesses the initial velocity
$\vert \vec{v}_0 \vert=1$. Obviously, we encounter Fermi
acceleration for all ensembles, i.e. driving amplitudes considered here. After an initial transient behavior of
typically $O(10^4)$ collisions characterized by relatively small variations of
the mean velocity, the function $\langle \vert \vec{v} \vert \rangle(n)$ becomes
an increasing power-law $\langle \vert \vec{v} \vert \rangle(n) \sim
n^{\beta(C)}$ ($\beta(C) > 0$).
The associated exponent $\beta$, as displayed in Table I, depends monotonically
on $C$. The diffusion in velocity space is anomalous in agreement
with the
fact that the dynamics in $F$-space is intermittent. This statement holds, of course, not only for $\langle \vert \vec{v} \vert \rangle(n)$ but also for the dependence on the real time  $\langle \vert \vec{v} \vert \rangle(t)$.
\begin{table}
\caption{Exponent $\beta(C)$ for $\langle \vert \vec{v} \vert \rangle(n) \sim
n^{\beta(C)}$ for different values of the amplitude $C$.}
\begin{ruledtabular}
\begin{tabular}{lccccccc}\label{tb:tab1}
$C$ & $0.05$ & $0.10$ & $0.20$ & $0.30$ & $0.40$ & $0.50$
&$0.60$\\
   $\beta (C)$ & $0.17$ & $0.28$ & $0.34$ & $0.36$ &$0.39$ & $0.40$ & $0.41$ \\
\end{tabular}
\end{ruledtabular}
\end{table}
The above analysis allows us to identify the basic ingredients of the
mechanism
responsible for Fermi acceleration in the driven elliptical
billiard. For $C=0$ the voluminous librator region in phase space contains a continuous set of
invariant and impenetrable manifolds. In the presence of the driving ($C \neq 0$) it becomes
penetrable and individual trajectories starting in the librator part of phase space are `pushed'
towards the separatrix (ejection phase).  Due to the driving the neighborhood of the separatrix is
replaced by a chaotic layer whose properties
depend on the amplitude $C$ of the boundary oscillation. While the dynamics is laminar in the
ejection phase, the motion in the chaotic layer is to a large
extent stochastic, introducing large  fluctuations of $F$ and consequently of the velocity  which in
turn leads to
Fermi acceleration.

In conclusion we have investigated the development of Fermi acceleration (FA) in the
harmonically driven elliptical billiard. Opposite to the
expectations according to the state-of-the-art of the field, we discover FA of the form $\langle \vert \vec{v}
\vert \rangle(n) \sim n^{\beta}$ with the exponent $\beta$ being tunable via the
amplitude of the oscillation of the elliptic boundary. We remark that no FA has been observed in
Ref. \cite{Koiller:1996} in the driven ellipse.  However,  Ref. \cite{Koiller:1996} employs a
slightly different driving mode. Additionally, the transient in which there is no FA is rather long
and could be easily misinterpreted as the absence of FA. Here we demonstrate for the first time that
FA is observed in a higher dimensional driven system for
which the corresponding static counterpart is integrable. To our knowledge, the considered system is also
the first billiard system showing an  acceleration law that is \textit{tunable} in a
straightforward  and controllable manner despite the complex structure of the underlying higher dimensional
phase space.  We show that the observed tunable
anomalous
diffusion in velocity space is associated with an intermittent dynamics of the second integral ($F$)
of the static integrable system, all by all triggered by the driving.

This work was supported by the German Research
Foundation (DFG) under the contract Schm885/13-1
and within the framework of the Excellence Initiative
through the Heidelberg Graduate School of Fundamental
Physics (GSC 129/1). F.L. acknowledges support from
the Landesgraduiertenf\"orderung Baden-W\"urttemberg. F.K.D. likes to thank the DAAD for financial support.

\end{document}